\documentclass{aa}

\usepackage[varg]{txfonts}
\usepackage{color}
\usepackage{graphicx}   
\usepackage{dcolumn}
\usepackage{bm}
\usepackage{dblfloatfix}
\usepackage{amsmath}    

\newcommand\simlt{\lower.5ex\hbox{$\; \buildrel < \over \sim \;$}}
\newcommand\simgt{\lower.5ex\hbox{$\; \buildrel > \over \sim \;$}}

\usepackage[colorlinks,citecolor=blue]{hyperref}

\begin{document}

\title{Analytic solution of a magnetized tori with magnetic polarization around Kerr black holes}
\author{Oscar M. Pimentel, F. D. Lora-Clavijo \and Guillermo A. Gonzalez}
\institute{Grupo de Investigaci\'on en Relatividad y Gravitaci\'on, Escuela de F\'isica, Universidad Industrial de Santander, A. A. 678, Bucaramanga 680002, Colombia}

\date{Received / Accepted}

\abstract{

We present the first family of magnetically polarized equilibrium tori around a Kerr black hole. The models were obtained in the test fluid approximation by assuming that the tori is a linear media, making it is possible to characterize the magnetic polarization of the fluid through the magnetic susceptibility $\chi_{m}$. The magnetohydrodynamic (MHD) structure of the models was solved by following the Komissarov approach, but with the aim of including the magnetic polarization of the fluid, the integrability condition for the magnetic counterpart was modified. We build two kinds of magnetized tori depending on whether the magnetic susceptibility is constant in space or not. In the models with constant $\chi_{m}$, we find that the paramagnetic tori ($\chi_{m}>0$) are more dense and less magnetized than the diamagnetic ones ($\chi_{m}<0$) in the region between the inner edge, $r_{in}$, and the center of the disk, $r_{c}$; however, we find the opposite behavior for $r>r_{c}$. Now, in the models with non-constant $\chi_{m}$, the tori become more magnetized than the Komissarov solution in the region where $\partial\chi_{m}/\partial r<0$, and less magnetized when $\partial\chi_{m}/\partial r>0$. Nevertheless, it is worth mentioning that in all solutions presented in this paper the magnetic pressure is greater than the hydrodynamic pressure. These new equilibrium tori can be useful for studying the accretion of a magnetic media onto a rotating black hole.

}


\titlerunning{Analytic solution of a magnetized tori...}
\authorrunning{Oscar M. Pimentel, F. D. Lora-Clavijo \& Guillermo A. Gonzalez}

\maketitle

\section{Introduction}
 
The accretion of a fluid onto a rotating black hole is believed to be the most powerful X-ray source in the universe, causing it to be an active research topic in astrophysics \citep{2002apa..book.....F}. Now, due to the strong gravitational field of the central object, it is necessary to consider a general relativistic approach to properly describe the dynamics of the fluid in the vicinity of the black hole. Additionally, one of the basic features in many theoretical models and numerical simulations of accretion disks is the presence of a strong magnetic field. This field has turned out to be very important because it interacts with the differential rotation of the disk and generates turbulence via the magneto-rotational instability \citep{1991ApJ...376..214B}. This turbulence provides the necessary viscous stress to transfer angular momentum, dissipate energy, and therefore generate the accretion process \citep{1998RvMP...70....1B}. Moreover, it is commonly assumed that strong poloidal magnetic fields are necessary to collimate and accelerate relativistic jets \citep{1977MNRAS.179..433B}. However, the origin and strength of those fields in the accretion disks are still a subject of study. Recent research has shown that a magnetically supported disk may be obtained from a thermally unstable hot disk \citep{2009ApJ...693..771F}. Another possibility is that the magnetic field increases gradually while it is dragged from the outer region to the inner region, near the black hole \citep{2012MNRAS.423.3083M}. 

It is well known that all substances contain spinning electrons that orbit around the nucleus, meaning we can model the microscopic structure of matter as an assembly of small dipoles, characterized macroscopically by the magnetization vector \citep{citeulike:4033945}. This vector enters in the Maxwell equations as a source of magnetic field, and could be important to understand the problems concerning the origin of the magnetic fields in astrophysical scenarios. Magnetization has already been considered for studying the equilibrium structure of neutron stars. For instance, in \cite{1982JPhC...15.6233B} the authors compute the magnetic susceptibility of the star crust and find that magnetization does not considerably change the surface properties, but may be connected to observable effects. Indeed, \cite{2010ApJ...717..843S} consider the possibility that the soft gamma-ray repeaters and the anomalous X-ray pulsars might be observational evidence for a diamagnetic phase transition that results in a domain formation. Furthermore, as mentioned by \cite{2016PASP..128j4201W}, neutron stars are also important for testing the Haas–van Alphen effect in which the magnetic susceptibility oscillates when the applied magnetic field is increased \citep{de1930dependence}. 

In the theory of galactic systems, \cite{navarro2018general} presented a static and axially symmetric self-gravitating thin disk in which the magnetic field is generated by a magnetization vector that is normal to the disk plane, presents a maximum at the disk center, and goes to zero at infinity. In accretion disk theory the magnetic polarization of the fluid has not yet been considered. Nevertheless, the first advances in this direction were made by \cite{2018arXiv180602266P}, where we presented the theoretical and numerical background to describe the evolution of a magnetically polarized fluid in a gravitational field. Among some results of this work, we can mention that the propagation speed of the fastest waves in the one-dimensional (1D) Riemann problems is greater in diamagnetic materials than in paramagnetic ones, and that the magnetic field and the relativistic character of the flows may increase considerably with the magnetic susceptibility.

Analytical models of accretion disks are of great interest in numerical simulations because they are used as initial data to study the nonlinear evolution of the fluid \citep{2008LRR....11....7F, 2013LRR....16....1A}. Among the analytical models, the Polish doughnuts \citep{1978A&A....63..221A} are the most used in numerical simulations due to their simplicity and the low computational expense they require \citep{2013LRR....16....1A}. These equilibrium models consist of a non-self-gravitating barotropic tori orbiting around a Kerr black hole. Later, \cite{2006MNRAS.368..993K} followed the method of Abramowicz to compute a magnetized tori with a purely toroidal magnetic field. These models have recently been used by \cite{2017MNRAS.467.1838F} to show that a strong toroidal magnetic field cannot be maintained during the disk evolution. Moreover, \cite{2018MNRAS.475..108B} used the Komissarov models to simulate the interplay of the Papaloizou-Pringle instability and the magneto rotational instability. Additionally, it is worth mentioning that \cite{2017A&A...607A..68G}, following the Komissarov procedure, found new magnetized equilibrium tori with a nonconstant angular momentum distribution in the disk. 

The analytic solutions for magnetized tori presented in the literature do not take into account the contribution of the magnetic dipoles in the torus. We therefore generalize in this work the KomisSarov models by including the magnetic polarization of the matter. For this purpose, we consider the energy-momentum tensor for a magnetically polarized fluid that is given in \cite{Maugin:1978tu, 2010PhRvD..81d5015H, 2015MNRAS.447.3785C}, and follow the Komissarov approach \citep{2006MNRAS.368..993K} to solve the hydrodynamic structure of the tori. These new models may be useful as initial data to study the evolution of a magnetic media in the gravitational field of a Kerr black hole. The organization of this paper is the following: in Sect. \ref{sec:equations} we present the Euler equations that describe the equilibrium structure of a magnetically polarized tori endowed with a toroidal magnetic field around a rotating black hole. To write these equations we concentrate on the special case in which the magnetization vector is parallel to the magnetic field, so that we can relate both vectors through the magnetic susceptibility, $\chi_{m}$. In Sect. \ref{sec:grpic} we present the conditions to write the Euler equations as an exact differential; then, as in \cite{2006MNRAS.368..993K}, we assume a barotropic tori in which the angular velocity is a function of the specific angular momentum only. However, the condition related to the magnetic counterpart is modified in order to include the magnetic polarization of the fluid. In Sect. \ref{sec:results} we present two kinds of magnetized tori: when the magnetic susceptibility is constant in space (Sect. \ref{subsec:case1}), and when it changes with coordinates (Sect. \ref{subsec:case2}). Finally, the main results are presented in Sect. \ref{sec:conclusions}. In this paper we use the signature ($-,+,+,+$) and geometrized units, for which $G=c=1$. 

\section{\label{sec:equations}Magnetohydrodynamic equations with magnetic polarization}

The dynamics of an ideal fluid with magnetic polarization in a magnetic field is described by the conservation laws
\begin{eqnarray}
\nabla_{\mu}T^{\mu\nu}=0,\label{momentum_conservation}\\
\nabla_{\mu}(\rho u^{\mu})=0,\label{mass_conservation}
\end{eqnarray}  
and by the relevant Maxwell equations,
\begin{equation}
\nabla_{\mu}(u^{\mu}b^{\nu}-b^{\mu}u^{\nu})=0,
\label{relevant_maxwell}
\end{equation}
where $\rho$ is the rest mass density and $b^{\mu}$ is the magnetic field, both measured in a reference frame that moves with the same four-velocity as the fluid, $u^{\mu}$. The magnetic polarization can then be characterized macroscopically through the magnetization vector $m^{\mu}$, which is defined as the magnetic dipole moment per unit volume. From now on, we concentrate on the physically important case in which $m^{\mu}$ and $b^{\mu}$ are related by means of the linear constitutive equation, $m^{\mu}=\chi b^{\mu}$, where $\chi=\chi_m/(1+\chi_m)$, and $\chi_m$ is the magnetic susceptibility. When $\chi_m <0$ the fluid is diamagnetic and when $\chi_m >0$ the fluid is paramagnetic. In the first case the polarization results from induced orbital dipole moments in a magnetic field \citep{griffiths2005introduction}, and in the second case the polarization is generated by magnetic torques in substances whose atoms have a nonzero spin dipole moment. The total energy-momentum tensor, $T^{\alpha\beta}$, for a magnetically polarized fluid was computed in \cite{Maugin:1978tu} and more recently in \cite{2015MNRAS.447.3785C} by following a different approach. The resulting tensor takes the form
\begin{eqnarray}
T^{\mu\nu}&=&\left[w+b^{2}(1-\chi)\right]u^{\mu}u^{\nu}+\left[p+\frac{1}{2}b^{2}(1-2\chi)\right]g^{\mu\nu}\nonumber\\
&&-(1-\chi)b^{\mu}b^{\nu},
\label{tensor_general}
\end{eqnarray}
in the linear media approximation. Here, $w$ is the enthalpy density, $p$ is the thermodynamic pressure, $b^{2}=b^{\mu}b_{\mu}$, and $g^{\mu\nu}$ is the metric tensor.

Following previous works \citep{2006MNRAS.368..993K, 2015MNRAS.447.3593W, 2017A&A...607A..68G}, we now assume the test fluid approximation, and the gravitational field as given by the Kerr metric in Boyer-Lindquist coordinates $(t,\phi,r,\theta)$. We also consider that the fluid is axisymmetric and stationary, so the physical variables do not depend either on the azimuthal angle $\phi$ or on the time $t$. Finally, we restrict the movement of the fluid in such a way that $u^{r}=u^{\theta}=0$, and we restrict the magnetic field topology to a purely toroidal one, so $b^{r}=b^{\theta}=0$. With these assumptions, the baryon number conservation (\ref{mass_conservation}) and the relevant Maxwell equations (\ref{relevant_maxwell}) are identically satisfied, and the equilibrium structure of the tori is obtained from the Euler equations $h_{~\nu}^{\gamma}\nabla_{\mu}T^{\mu\nu}=0$, where $h_{~\nu}^{\gamma}=\delta^{\gamma}_{~\nu}+u^{\gamma}u_{\nu}$ is the projection tensor. This contraction leads to the following expression.
\begin{eqnarray}
\left[w+(1-\chi)b^{2}\right]u_{\mu}u^{\mu}_{~,i}&+&\left[p+\frac{1}{2}(1-2\chi)b^{2}\right]_{,i}\nonumber\\&-&(1-\chi)b_{\mu}b^{\mu}_{~,i}+\frac{1}{2}(1-\chi)b^{2}_{~,i}=0,
\label{eulerv1}
\end{eqnarray}
where $i=r,\theta$. 

Nevertheless, it is useful to write this last equation in terms of the angular velocity
\begin{equation}
\Omega=\frac{u^{\phi}}{u^{t}}=-\frac{g_{\phi t}+lg_{tt}}{g_{\phi\phi}+lg_{t\phi}},
\label{omega}
\end{equation}
and the specific angular momentum
\begin{equation}
l=-\frac{u_{\phi}}{u_{t}}=-\frac{g_{\phi t}+\Omega g_{\phi\phi}}{g_{tt}+\Omega g_{t\phi}},
\label{angular}
\end{equation}
in such a way that the Euler equations take the form
\begin{equation}
(\ln|u_{t}|)_{,i}-\frac{\Omega}{1-l\Omega}l_{,i}+\frac{p_{,i}}{w}-\frac{(\chi p_m)_{,i}}{w}+\frac{[(1-\chi)\mathcal{L} p_m]_{,i}}{\mathcal{L}w}=0,
\label{eulerv2}
\end{equation}
where $p_{m}=b^{2}/2$ and $\mathcal{L}=g_{t\phi}^{2}-g_{tt}g_{\phi\phi}$. It is important to mention that when $\chi=0$, Eq. (\ref{eulerv2}) reduces to Eq. (14) in \cite{2006MNRAS.368..993K}, where the tori do not have magnetic polarization.

\section{\label{sec:grpic} Integrability conditions}

In order to write the first three terms in Eq. (\ref{eulerv2}) as an exact differential, we follow the procedure used by \cite{2006MNRAS.368..993K} where it is supposed that the fluid obeys a barotropic equation of state $w=w(p)$, and that the surfaces of the $\Omega$ and $l$ constants coincide in such a way that $\Omega=\Omega(l)$ \citep{1978A&A....63..221A}. With these two assumptions, Eq. (\ref{eulerv2}) takes the form 
\begin{equation}
d\left(\ln|u_{t}|+\int_{0}^{p}\frac{dp}{w}-\int_{0}^{l}\frac{\Omega dl}{1-l\Omega}\right)+\mathcal{I}=0,
\label{euler_exact}
\end{equation} 
where
\begin{eqnarray}
\mathcal{I}&=&-\frac{d(\chi p_m)}{w}+\frac{d[(1-\chi)\mathcal{L} p_m]}{\mathcal{L}w}\nonumber\\
&=&\frac{1-2\chi}{w}dp_{m}+(1-\chi)\frac{p_{m}}{\mathcal{L}w}d\mathcal{L}-\frac{2p_{m}}{w}d\chi.
\label{exact_diff}
\end{eqnarray}
We note that in this equation, we take $\chi$ to be an arbitrary function of the coordinates. Nevertheless, with the aim of writing $\mathcal{I}$ as a exact differential, we assume that $\chi=\chi(\mathcal{L})$. In this way, $d\chi=(\partial\chi/\partial\mathcal{L})d\mathcal{L}$, and the last equation reduces to
\begin{equation}
\mathcal{I}=\frac{1-2\chi}{w}dp_{m}+\left(1-\chi-2\chi'\mathcal{L}\right)\frac{p_m}{w\mathcal{L}}d\mathcal{L}
\label{exact_case2}
,\end{equation}
where $\chi'=\partial\chi/\partial\mathcal{L}$. Now, if we can find a function $z=z(p_{m},\mathcal{L})$ that satisfies the derivatives
\begin{equation}
\frac{\partial z}{\partial p_{m}}=\frac{1-2\chi}{w},\hspace{5mm}\frac{\partial z}{\partial\mathcal{L}}=\left(1-\chi-2\chi'\mathcal{L}\right)\frac{p_{m}}{w\mathcal{L}},
\label{derivative}
\end{equation}
such that $\mathcal{I}=(\partial z/\partial p_{m})dp_{m}+(\partial z/\partial \mathcal{L})d\mathcal{L}$, then Eq. (\ref{exact_case2}) could also be written as an exact differential $\mathcal{I}=dz$. 

By integrating Eq. (\ref{derivative}) we obtain the general solution
\begin{equation}
z=(1-2\chi)\int_{p_{m_{0}}}^{p_{m}}\frac{dp_{m}}{w}+p_{m_{0}}\left.\int_{\mathcal{L}_{0}}^{\mathcal{L}}\frac{1-\chi-2\chi'\mathcal{L}}{w\mathcal{L}}d\mathcal{L} ~ \right|_{p_{m_{0}}}+C_{0},
\label{z}
\end{equation}
 if the condition
\begin{equation}
z-p_{m}\int_{\mathcal{L}_{0}}^{\mathcal{L}}\frac{1-\chi-2\chi'\mathcal{L}}{\mathcal{L}w}d\mathcal{L}=\mathcal{F}(p_{m})
\label{integrability}
\end{equation} 
is satisfied, $C_{0}$  being a constant of integration, and $\mathcal{F}(p_{m})$ a function of the magnetic pressure only. Thus, the problem reduces to find a relation between $w$, $p_{m}$, and $\mathcal{L}$ that satisfies Eq. (\ref{integrability}). Fortunately, it is possible to transform this condition into a partial differential equation for the enthalpy by differentiating it, first with respect to $p_m$, and then with respect to $\mathcal{L}$, 
\begin{equation}
(1-2\chi)\mathcal{L}\frac{\partial w}{\partial \mathcal{L}}-p_{m}(1-\chi-2\chi'\mathcal{L})\frac{\partial w}{\partial p_{m}}+(1-\chi)w=0.
\label{general_w_solution}
\end{equation}
We solve this equation with the method of the characteristics and find that any function of the form
\begin{equation}
w=\mathcal{L}^{-1/2}e^{-\phi}f[(1-2\chi)p_{m}\mathcal{L}^{1/2}e^{\phi}],
\label{general_w_solution}
\end{equation}
is a solution to the equation. In this last expression,
\begin{equation}
\phi=\int_{\mathcal{L}_{0}}^{\mathcal{L}}\frac{d\mathcal{L}}{2\mathcal{L}(1-2\chi)},
\label{phi}
\end{equation} 
and $f$ is an arbitrary function of its argument, so we can obtain different solutions depending on the particular choice of $f$. 

In order to find magnetically polarized tori that reduce to the Komissarov solution when $\chi=0$, we now assume $f$ as the function
\begin{equation}
f=\tilde{K}_{m}\left[(1-2\chi)p_{m}\mathcal{L}^{1/2}e^{\phi}\right]^{1/\eta},
\label{f_fun}
\end{equation}
where $\tilde{K}_{m}$ and $\eta$ are arbitrary constants. We can then obtain the magnetic pressure from Eq. (\ref{general_w_solution}) in such a way that 
\begin{equation}
p_{m}=\frac{K_{m}}{1-2\chi}\mathcal{L}^{(\eta-1)/2}e^{(\eta-1)\phi}w^{\eta},
\label{mag_pres_caso2}
\end{equation}
with $K_{m}=\tilde{K}_{m}^{~-\eta}$. Additionally, we need to choose a particular relation between $\chi$ and $\mathcal{L}$ to compute $\phi$ from (\ref{phi}) and completely determine $p_{m}$. Subsequently, with the aim of analyzing disks with different magnetic polarization states, we adopt the form
\begin{equation}
\chi=\chi_{0}+\chi_{1}\mathcal{L}^{\alpha},
\label{chi}
\end{equation}
where $\chi_{0}$, $\chi_{1}$, and $\alpha$ are constants. Therefore, depending on the choice of these constants we obtain a different magnetic susceptibility. When $\chi_{1}=0,$ this susceptibility is constant, when $\chi_{1}\neq 0$ it becomes a function of the spatial coordinates, and when $\chi_{1}=\chi_{0}=0$ it is zero and then we reduce to the Komissarov solution. 

Once the function $\chi$ has been defined, the magnetic pressure in Eq. (\ref{mag_pres_caso2}) takes the simple form
\begin{equation}
p_{m}=K_{m}\mathcal{L}^{\tilde{\lambda}}w^{\eta}\tilde{f},
\label{mag_pres}
\end{equation}
with,
\begin{equation}
\tilde{\lambda}=\frac{1-\chi_{0}}{1-2\chi_0}(\eta-1),\hspace{5mm}\tilde{f}=(1-2\chi)^{\frac{1-\eta}{2\alpha(1-2\chi_{0})}-1}.
\label{functions_case2}
\end{equation}
Finally, with the enthalpy computed from Eq. (\ref{mag_pres}), we solve the integrals in (\ref{z}) to obtain
\begin{equation}
z=(1-2\chi)\frac{\eta}{\eta-1}\frac{p_{m}}{w},
\label{z2_caso2}
\end{equation}
as a particular solution of Eq. (\ref{derivative}) in which $\chi$ is given by Eq. (\ref{chi}).

With the function $z(p_{m},\mathcal{L})$ that allows us to write $\mathcal{I}=dz$, the Euler equations in (\ref{euler_exact}) can be solved in the form
\begin{equation}
\ln|u_{t}|+\int_{0}^{p}\frac{dp}{w}-\int_{0}^{l}\frac{\Omega dl}{1-l\Omega}+(1-2\chi)\frac{\eta}{\eta-1}\frac{p_{m}}{w}=\text{constant},
\label{euler_exact_final}
\end{equation} 
where the integrals are computed, as in \cite{2006MNRAS.368..993K},  by assuming a torus with constant angular momentum $l=l_0$, and a polytropic equation of state for the gas pressure, 
\begin{equation}
p=Kw^{\kappa},
\label{polytropic_EOS}
\end{equation}
where $K$ and $\kappa$ are constants. Therefore, Eq. (\ref{euler_exact_final}) becomes
\begin{equation}
\ln|u_{t}|+\frac{\kappa}{\kappa-1}\frac{p}{w}+(1-2\chi)\frac{\eta}{\eta-1}\frac{p_{m}}{w}=\text{constant}.
\label{euler_exact_final_chi}
\end{equation}
On the other hand, it is usual to introduce the relativistic effective potential (\cite{1978A&A....63..221A}),
\begin{equation}
W=\ln|u_{t}|+\int_{l}^{l_\infty}\frac{\Omega dl}{1-l\Omega},
\label{effective}
\end{equation}
which reduces to $W=\ln|u_{t}|$ in the case of a constant angular momentum torus. With $W$, the Euler equations in the form (\ref{euler_exact_final_chi}) can be expressed as 
\begin{equation}
W-W_{in}+\frac{\kappa}{\kappa-1}\frac{p}{w}+(1-2\chi)\frac{\eta}{\eta-1}\frac{p_{m}}{w}=0,
\label{euler1}
\end{equation}
with $W_{in}=\ln|u_{t_{in}}|$ being the potential defined at the inner edge of the disk, where in turn $p_{in}=0$ and $p_{m_{in}}=0$. It is worth mentioning that when the magnetic polarization of the fluid is zero ($\chi=0$), Eqs. (\ref{mag_pres}) and (\ref{euler1}) in this paper reduce to Eqs. (29) and (30) in \cite{2006MNRAS.368..993K}. Therefore, the new magnetized tori with magnetic polarization reduce to the Komissarov solution in the case where $\chi=0$.

\section{\label{sec:results}Results}

The new analytic solutions obtained in the previous sections have the following free parameters: the potential at the inner edge of the disk $W_{in}$, the angular momentum of the fluid $l_{0}$, the exponents of the pressures $\kappa$ and $\eta$, the magnetic polarization parameters $\chi_{0}$, $\chi_{1}$ and $\alpha$, and finally, the enthalpy and the magnetization parameter at the disk center, $w_{c}$ and $\beta_{c}=p_{c}/p_{m_{c}}$ , respectively. The disk center, $r_{c}$, is defined as one of the points where the gradient pressure vanishes, so the angular momentum $l_{0}$ equals the Keplerian angular momentum,
\begin{equation}
l_{k}=\pm\frac{r^{2}\mp 2ar^{1/2}+a^{2}}{r^{3/2}-2r^{1/2}\pm a},
\label{Keplerian}
\end{equation}  
where $a$ is the angular momentum of the black hole, and the upper signs correspond to prograde motion. We highlight the fact  that in these formulae we have chosen $M=1$, $M$  being the mass of the black hole. By doing $l_{k}=l_{0}$ we therefore compute $r_{c}$.

The procedure to compute the physical variables consists in finding the gas pressure and the magnetic pressure at the disk center through the equations
\begin{eqnarray}
p_{c}&=&w_{c}(W_{in}-W_{c})\left(\frac{\kappa}{\kappa-1}+\frac{\eta}{\eta-1}\frac{1-2\chi_{c}}{\beta_{c}}\right)^{-1},\\
p_{m_{c}}&=&\frac{p_{c}}{\beta_{c}},
\end{eqnarray}
from which we can compute the constants $K$ and $K_{m}$. Next, the enthalpy as a coordinate function can be found from Eq. (\ref{euler1}). With the enthalpy, the magnetic pressure is determined by Eq. (\ref{mag_pres}), and the gas pressure by Eq. (\ref{polytropic_EOS}). The magnetic field components can be obtained from the constraint $u^{\mu}b_{\mu}=0$, and the specific angular momentum $l=-u_{\phi}/u_{t}$, which lead to the following expressions.
\begin{equation}
b^{\phi}=\pm\sqrt{\frac{2p_{m}}{\mathcal{A}}}, \hspace{5mm} b^{t}=l_{0}b^{\phi},
\label{b_components}
\end{equation}
with $\mathcal{A}=g_{\phi\phi}+2l_{0}g_{t\phi}+l_{0}^{2}g_{tt}$. Finally, the four-velocity is computed from,
\begin{equation}
u^{t}=-\frac{1}{u_{t}(1-l_{0}\Omega)}, \hspace{5mm} u^{\phi}=\Omega u^{t},
\label{u_components}
\end{equation}  
where $(u_{t})^{2}=\mathcal{L}/\mathcal{A}$.

Since we are interested in analyzing the contribution of the magnetic polarization to the physical variables of the fluid, we now compare our new disk solutions with the model A proposed by \cite{2006MNRAS.368..993K}, in which the disk has a constant angular momentum of $l_{0}=2.8$ and an inner-edge effective potential of $W_{in}=-0.030$. These parameters, along with the black hole's spin parameter $a=0.9$, correspond to a disk with finite outer radius, whose center is located at $r_{c}=4.62$. Finally, in the Komissarov model $\kappa=\eta=4/3$, $\beta_{c}=0.1$, and $w_{c}=1$. The three remaining parameters $\chi_{0}$, $\chi_{1}$, and $\alpha$ determine the magnetic polarization state of the disk, and are analyzed in the following two sections. The first of these sections is dedicated to magnetized tori with constant magnetic susceptibility, while in the second section we discuss the effect of a nonconstant $\chi_{m}$ on the equilibrium state of the disks. 

\subsection{\label{subsec:case1} Magnetized disks with constant magnetic susceptibility}

In many cases, it is reasonable to consider that the spatial variations of the magnetic susceptibility are very small, so we can determine the magnetic polarization of a material through a single constant. In our models, a disk with such an approximation can be obtained by taking $\chi_{1}=0$, which in turn implies that $\chi=\chi_{0}$. We construct two diamagnetic disks with magnetic susceptibilities of $\chi_{m}=-0.2, -0.4$, which correspond to $\chi_{0}\approx-0.25,-0.67$, and two paramagnetic disks with $\chi_{m}= 0.2, 0.4$, which imply that $\chi_{0}\approx 0.17, 0.29$. These models will be compared with the Komissarov solution, which corresponds to the case $\chi_{0}=0$ ($\chi_{m}=0$). 

\begin{figure*}
\begin{center}
\begin{tabular}{ccc}
\includegraphics[height=2.1in]{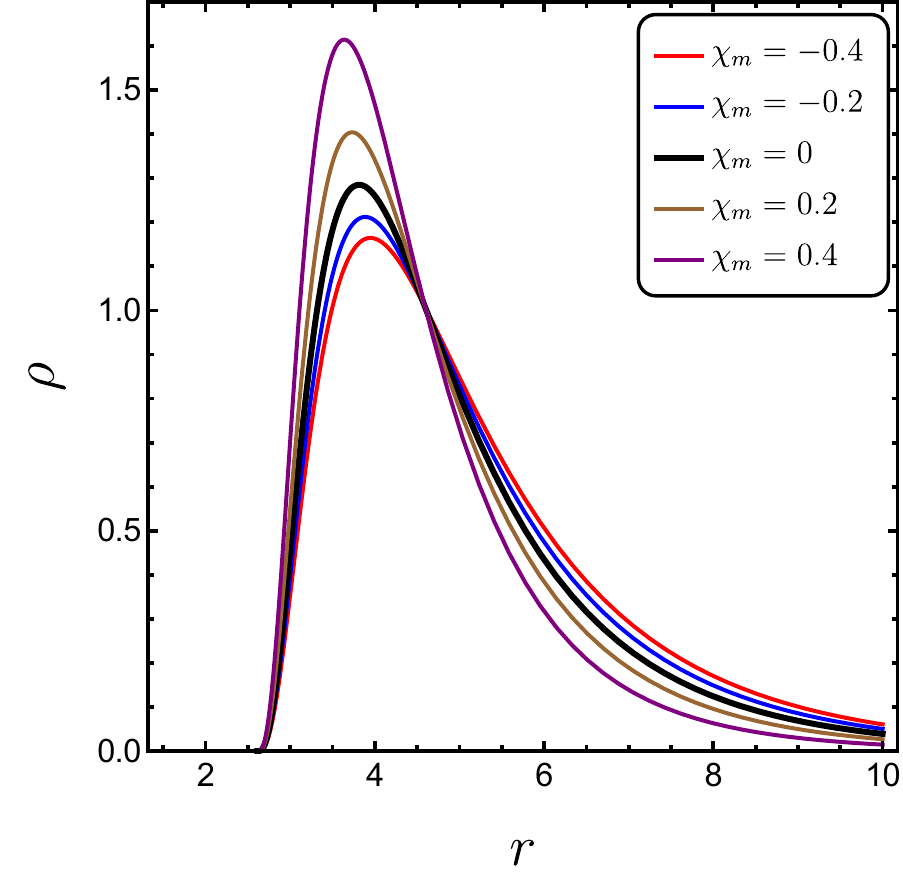} & \includegraphics[height=2.1in]{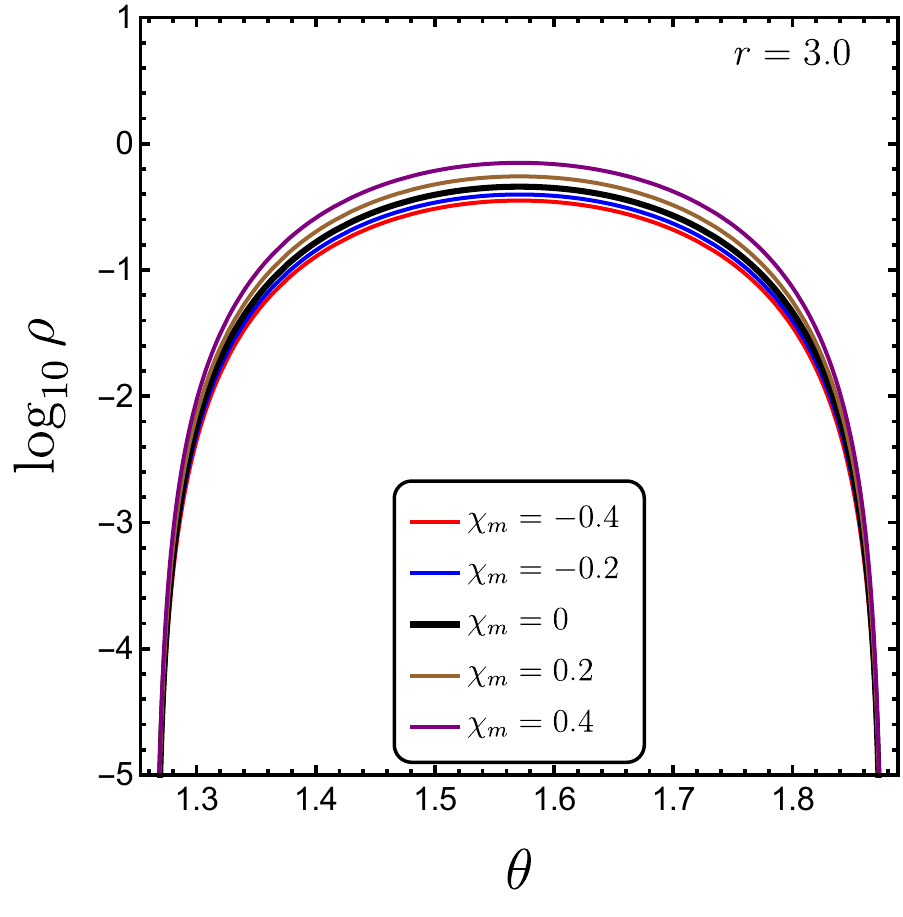} & \includegraphics[height=2.1in]{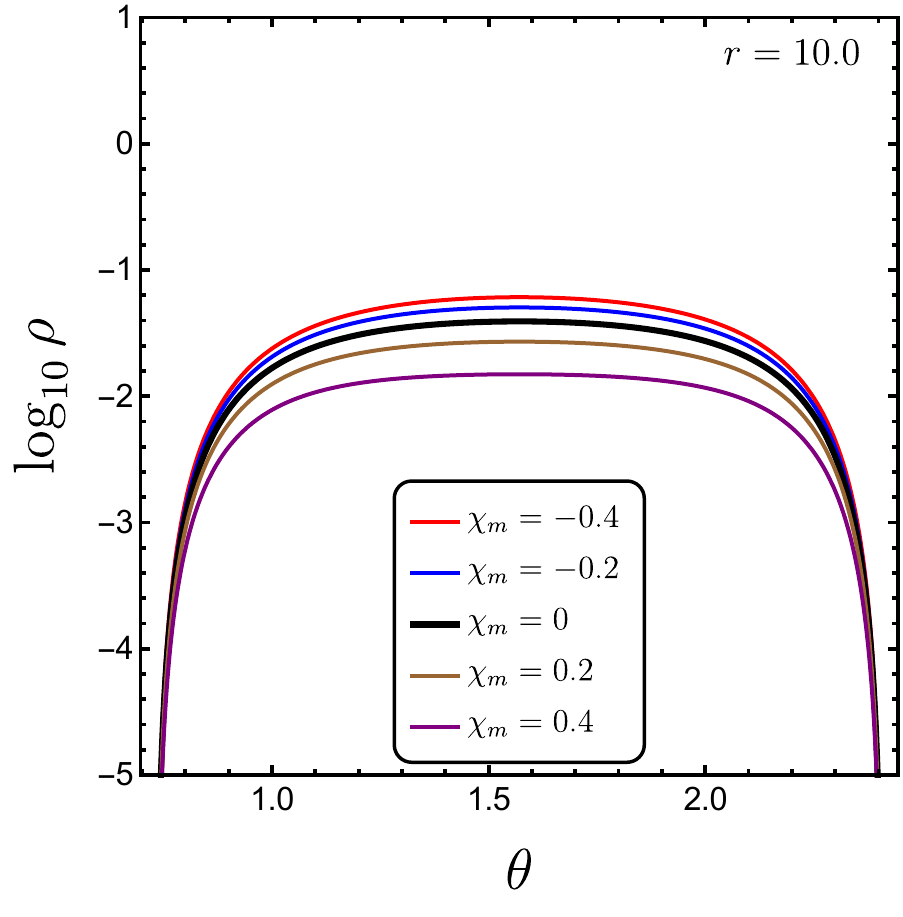}
\end{tabular}
\end{center}
\begin{center}
\begin{tabular}{ccc}
\includegraphics[height=2.0in]{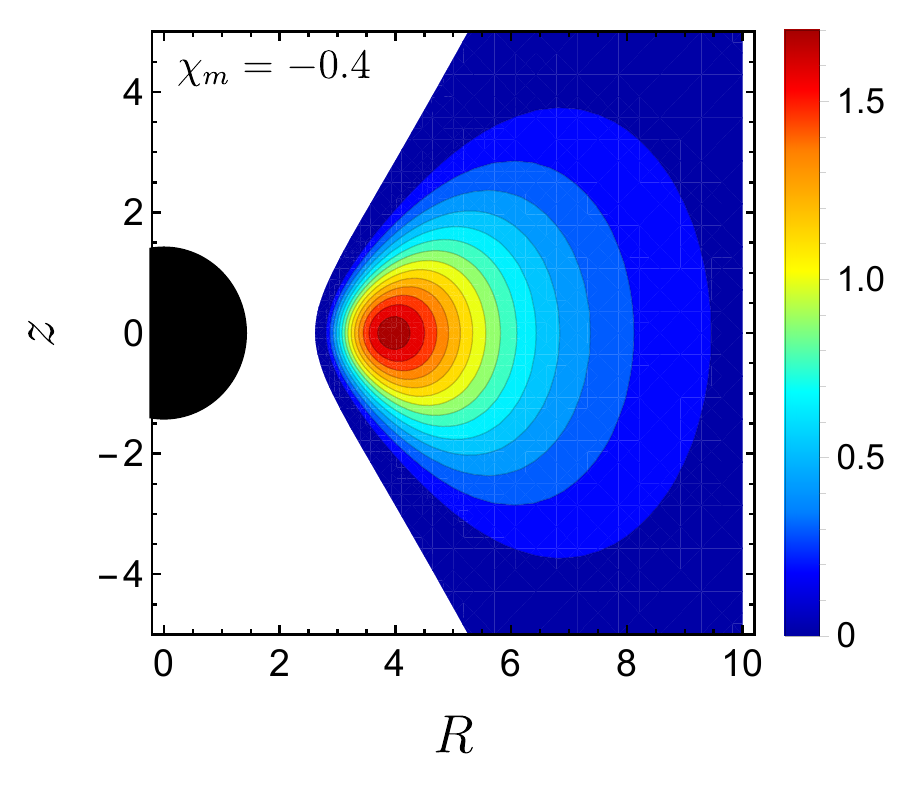} & \includegraphics[height=2.0in]{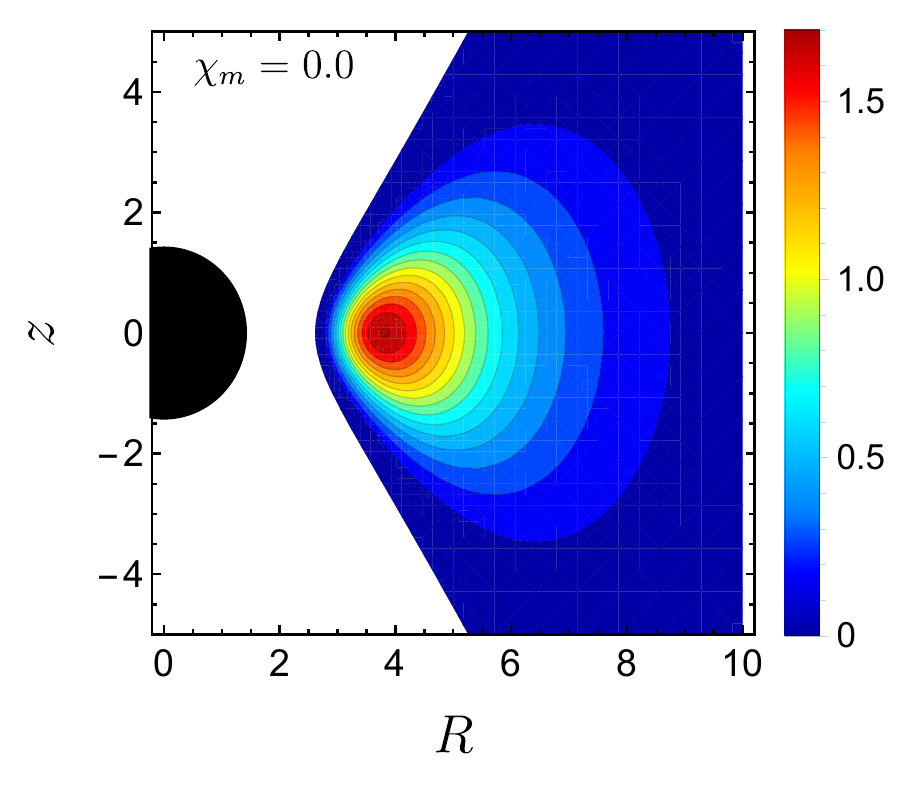} & \includegraphics[height=2.0in]{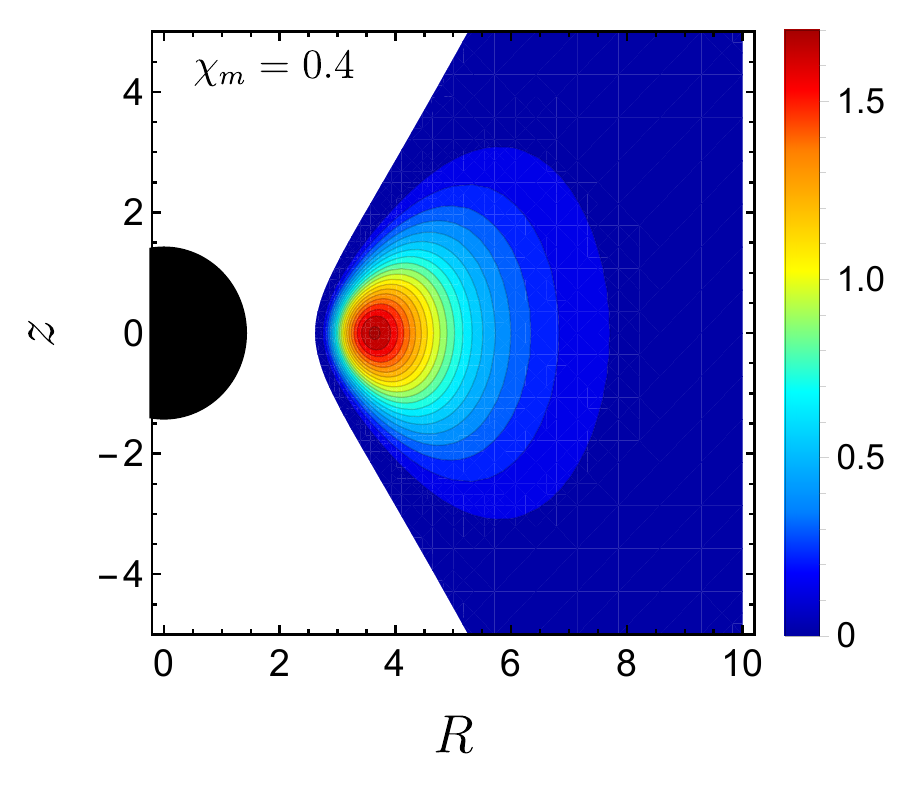}
\end{tabular}
\end{center}
\caption{Spatial distribution of the rest mass density $\rho$ for different values of magnetic susceptibility $\chi_{m}$. The top-left panel is the radial dependence of $\rho$ on the equatorial plane defined by $\theta=\pi/2$. The top-center and top-right panels correspond to the angular dependence of $\rho$ in the radial positions $r=3.0$ and $r=10.0$, respectively. In the top row the black curves correspond to the Komissarov solution, obtained by taking $\chi_{m}=0$. The bottom row shows the density distribution of $\rho$ in the $z$-$R$ plane, with $R=r\sin\theta$ and $z=r\cos\theta$, for $\chi_{m}=-0.4, 0.0, 0.4$.}
\label{variables_chi_cte}
\end{figure*}

In Fig. $\ref{variables_chi_cte}$ we present the spatial behavior of the rest mass density for different values of magnetic susceptibility. The top-left panel is the radial profile of $\rho$ in the equatorial plane $\theta=\pi/2$, while the top-center and top-right panels show the angular behavior of the same physical variable at $r=3.0$ (close to the inner edge of the diks) and $r=10$, respectively. In these three panels, the black curve corresponds to the Komissarov solution. The bottom row shows the spatial distribution of the rest mass density in the meridional plane of the disk when the fluid is diamagnetic with $\chi_{m}=-0.4$ and paramagnetic with $\chi_{m}=0.4$. We also plot the Komissarov disk ($\chi_{m}=0.0$) for comparison purposes. From this figure, we note that the magnetic susceptibility considerably changes the way in which the matter is distributed within the disk. More specifically, the paramagnetic tori are more dense than the tori without magnetic polarization in the region between the inner edge, $r_{in}$, and the center of the disk, $r_{c}$. On the contrary, the diamagnetic disks are less dense than the Komissarov tori in this region. Nevertheless, in $r>r_{c}$ the rest mass density is greater in the diamagnetic disks than in those with paramagnetic properties. Additionally, we can see from the contour plots that the paramagnetic tori are more compact than the Komissarov solution, and even more compact than the diamagnetic ones.

\begin{figure*}
\begin{center}
\begin{tabular}{ccc}
\includegraphics[height=2.1in]{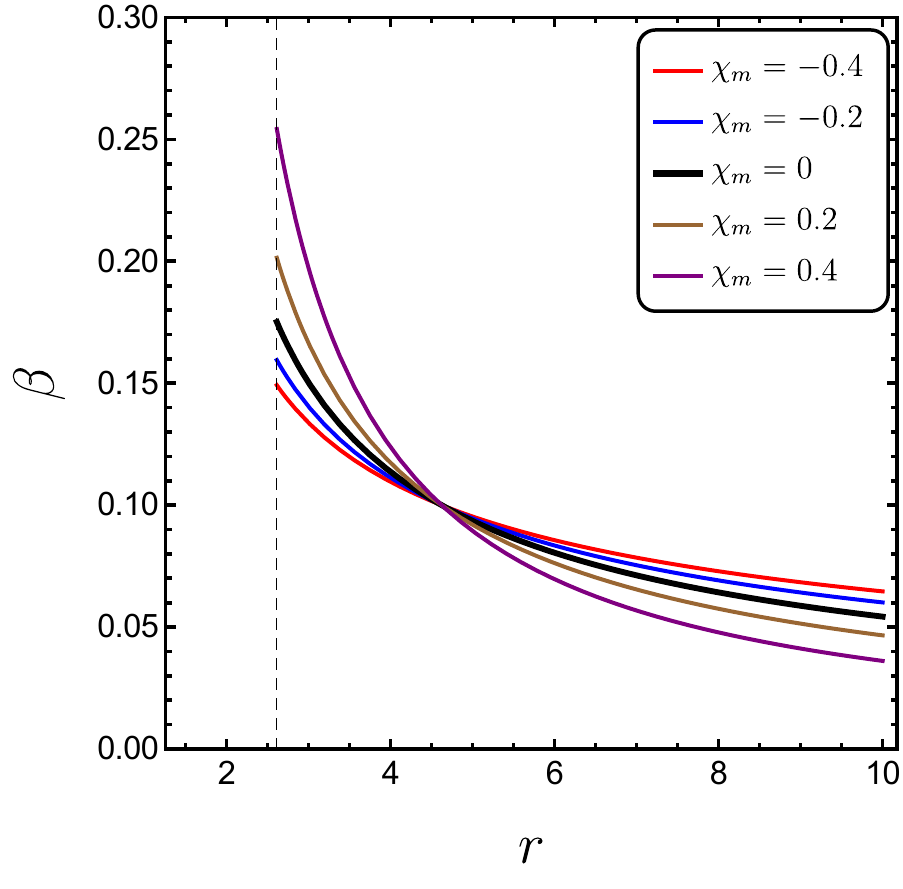} & \includegraphics[height=2.1in
]{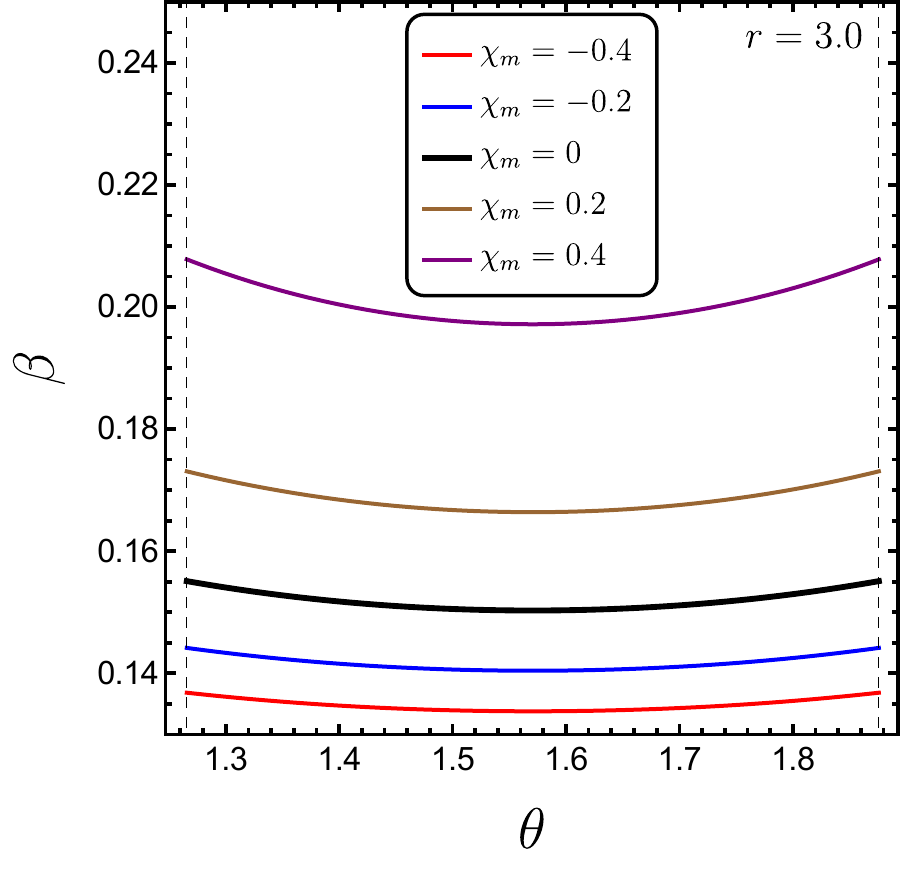} & \includegraphics[height=2.1in
]{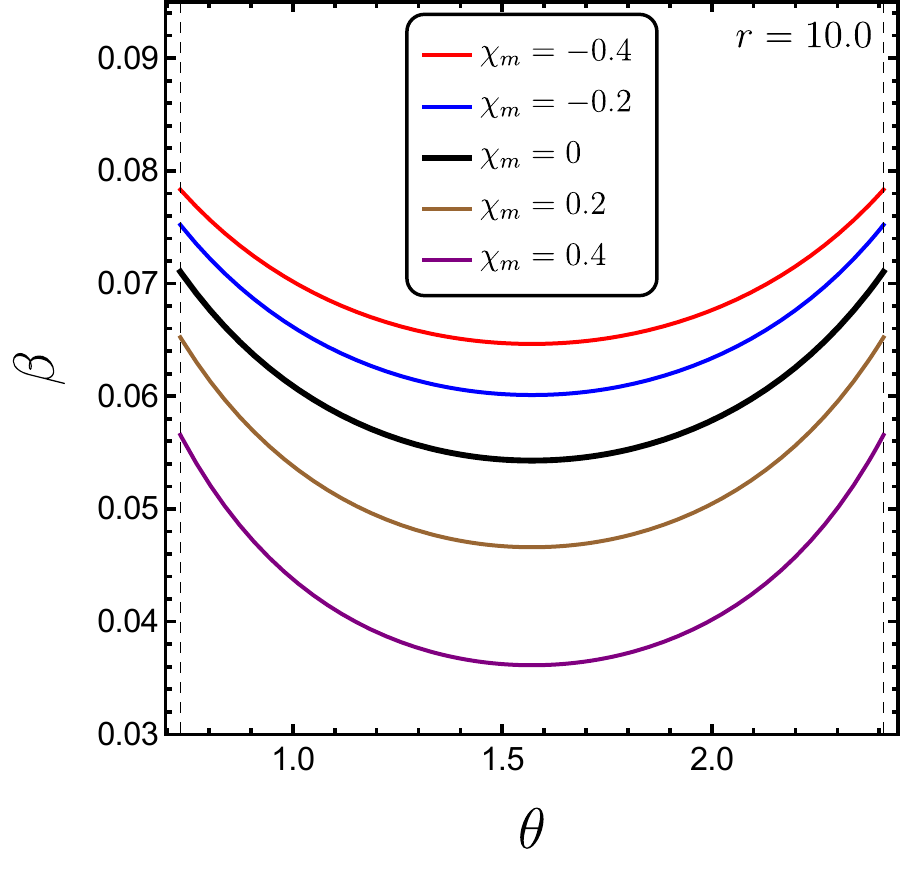}
\end{tabular}
\end{center}
\caption{Magnetization parameter $\beta$ as a function of the spatial coordinates for different values of magnetic susceptibility $\chi_{m}$. The left panel describes the radial dependence of $\beta$ on the equatorial plane ($\theta=\pi/2$). The middle and right panels show the angular dependence of $\beta$ at the position $r=3.0$ and $r=10.0$, respectively. The vertical dotted lines represent the boundaries of the disk.}
\label{beta}
\end{figure*}

On the other hand, in Fig. \ref{beta} we show the magnetization parameter $\beta$ for different values of magnetic susceptibility. The left panel describes the behavior of $\beta$ on the equatorial plane, while the center and right panels describe the magnetization parameter at $r=3.0$ and $r=10.0$, respectively. Additionally, the dotted vertical lines are the edges of the disk, whose positions can be computed from $W=W_{in}$. This figure shows that when the diamagnetic character of the disks increases (a more negative value of $\chi_{m}$), they become more magnetized in the region $r<r_{c}$ and less magnetized in $r>r_{c}$, as compared to the Komissarov solution. The paramagnetic tori exhibit the opposite behavior: they are less magnetized in $r<r_{c}$ and more magnetized in $r>r_{c}$ than the torus without magnetic polarization.

\subsection{\label{subsec:case2} Magnetized disks with nonconstant magnetic susceptibility}

As we mentioned at the end of Sect. \ref{sec:equations}, spatial variations of $\chi_{m}$ may affect the stationary state of the disk, so it is also interesting to consider a fluid with a nonconstant magnetic susceptibility. This kind of system has its motivation in some physical phenomena; for instance, the temperature dependence of $\chi_{m}$ in paramagnetic materials (\cite{gabold2018structural}), the rapid decrease of the magnetic susceptibility in a ferromagnetic-paramagnetic phase transition (\cite{1964JAP....35.2424A, 2002JPCM...14L.365C}), and the interaction between the spin of a particle and the gravitational field. This last effect is described by the Mathisson-Papapetrou-Dixon equations (\cite{1951RSPSA.209..248P, dixon1970dynamics}), and has been studied in the kerr spacetime, for instance in \cite{1998PhRvD..58f4005S} and \cite{1999MNRAS.308..863S}.

\begin{figure*}
\begin{center}
\begin{tabular}{ccc}
\includegraphics[height=2.0in]{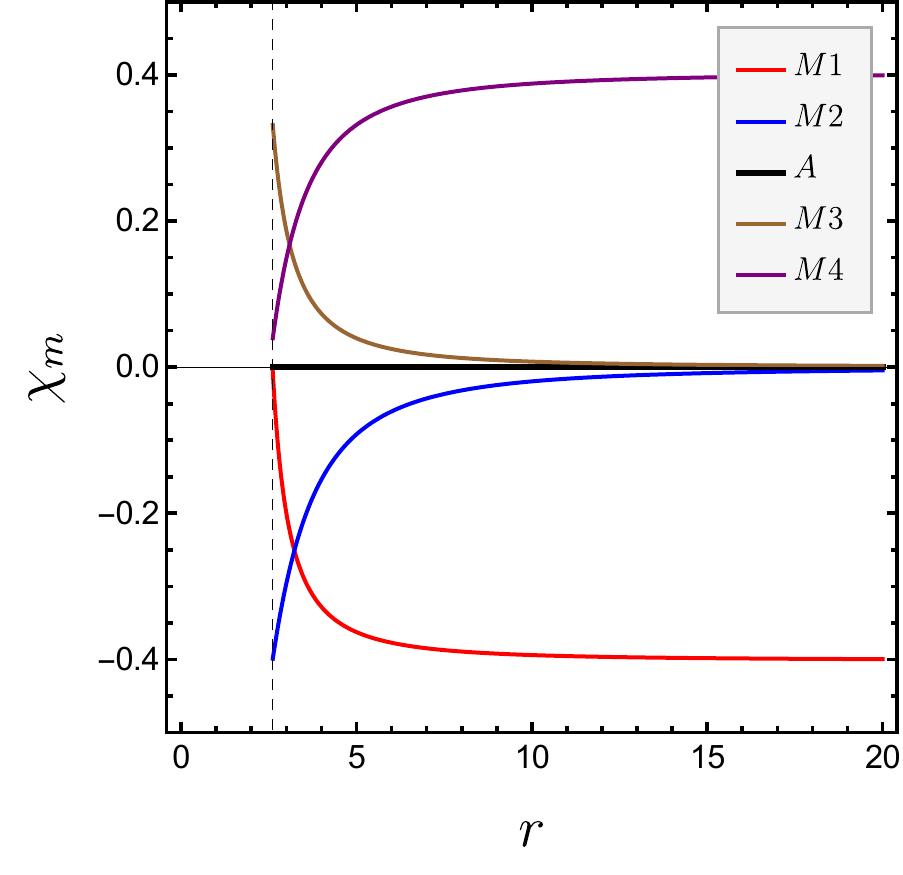} & \includegraphics[height=2.0in]{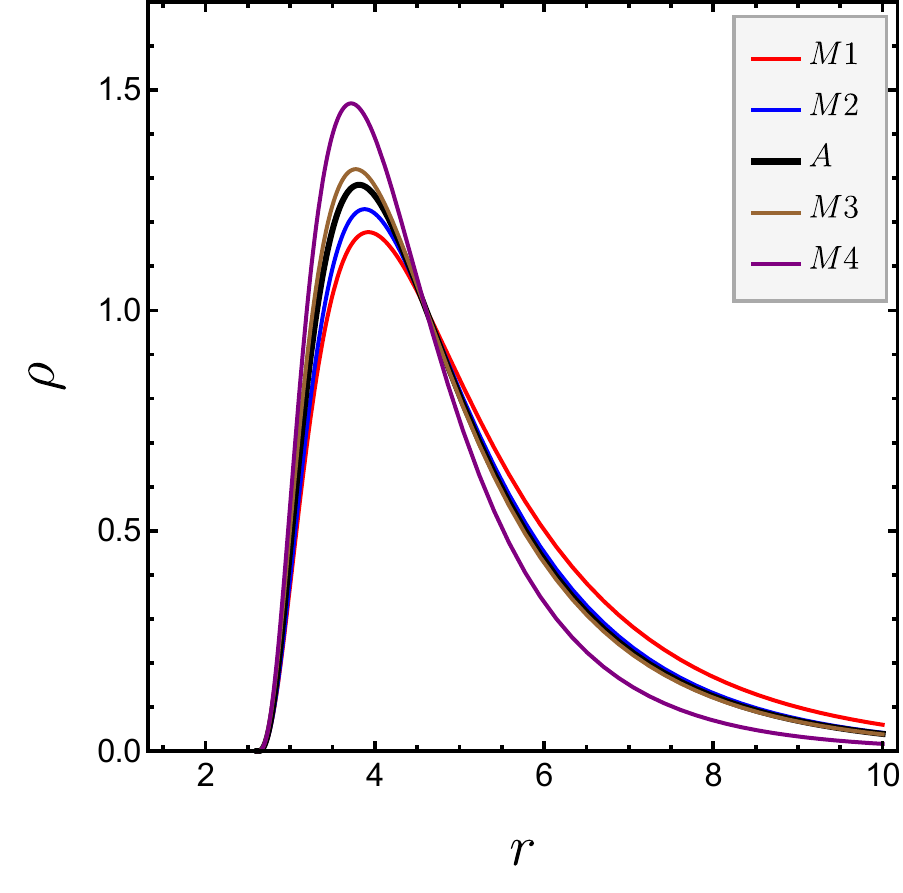} & \includegraphics[height=2.0in]{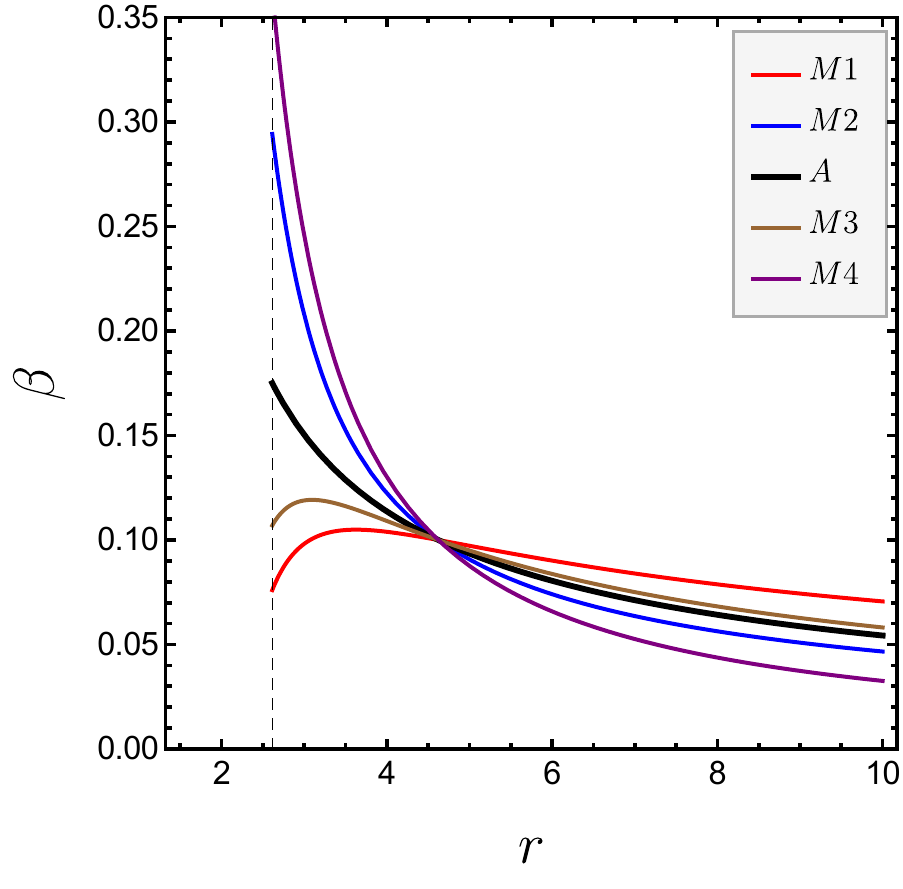}\\
\includegraphics[height=2.0in]{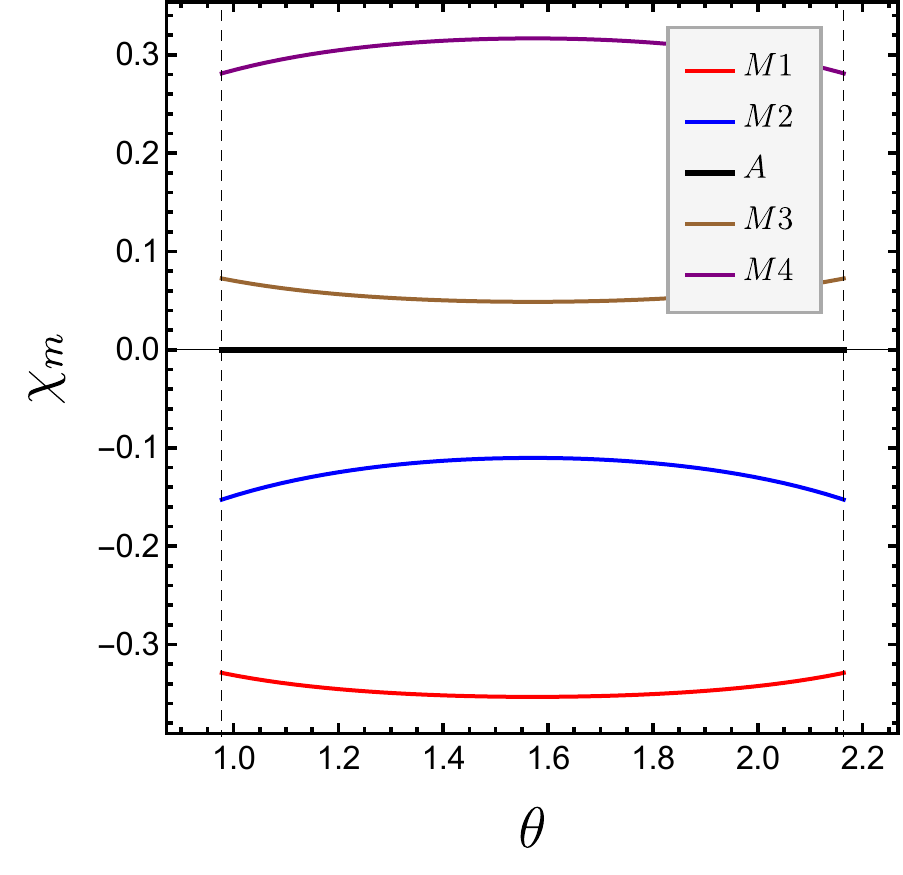} & \includegraphics[height=2.0in]{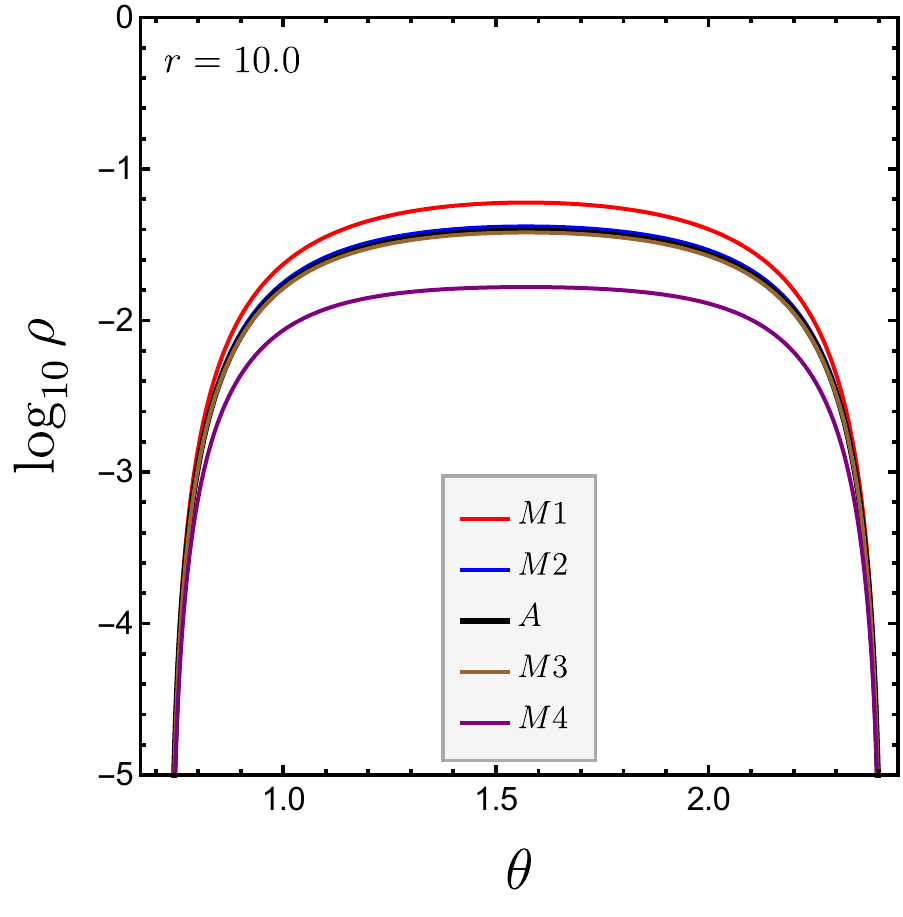} & \includegraphics[height=2.0in]{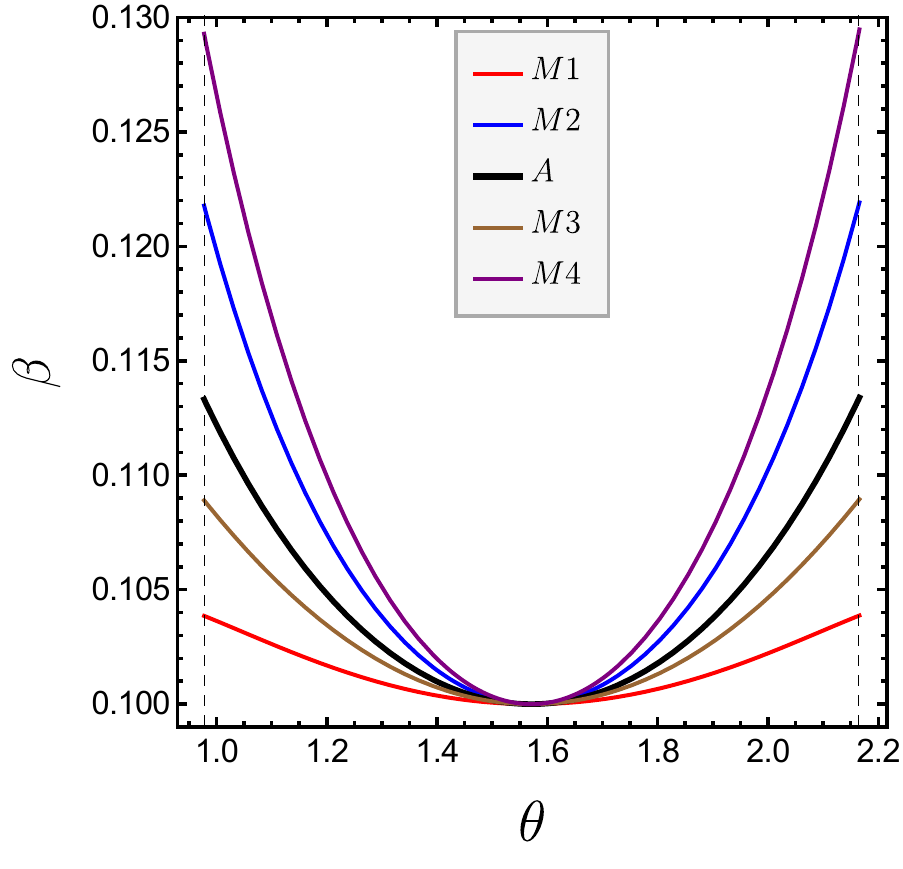}
\end{tabular}
\end{center}
\caption{Magnetic susceptibility $\chi_{m}$ (left column), rest mass density $\rho$ (middle column), and magnetization parameter $\beta$ (right column) as a function of the spatial coordinates ($r$, $\theta$) for the models M1, M2, M3, M4, and for the Komissarov solution, labeled by the letter $A$. Each one of these models is characterized by a different magnetic susceptibility function that is shown in the first column. The top row shows the radial behavior of $\chi_{m}$, $\rho$, and $\beta$ on the equatorial plane $\theta=\pi/2$, and the bottom row shows the angular dependence of the same physical variables at the center of the disk $r=r_c$, except for the mass density, for which the behavior is showed at $r=10.0$. The vertical dotted lines correspond to the boundaries of the disk.}
\label{variables_chi_NOcte}
\end{figure*}

With the aim of analyzing the stationary state of a fluid with a nonconstant magnetic susceptibility, we construct two diamagnetic disk models M1 and M2, and two paramagnetic ones M3 and M4. In Table \ref{table1} we present the parameters $\chi_{0}$, $\chi_{1}$, and $\alpha$ that define the magnetic susceptibility (\ref{chi}) in each model. The radial and angular behavior of $\chi_{m}$ in the disk is showed in the left column of Fig. \ref{variables_chi_NOcte}. We note that in the four models that we propose, the magnetic susceptibility changes considerably quickly with the radial coordinate, especially in the region between the inner edge and the center of the disk, while $\chi_m$ remains approximately constant in $r\gtrsim 10$. In the diamagnetic models M1 and M2, the rest mass density is reduced in the inner region of the disk ($r\lesssim 5$) and it is higher in the outer region, in comparison with the Komissarov solution, which is labeled with letter $A$ in Fig. \ref{variables_chi_NOcte}. On the contrary, the mass density in the paramagnetic models M3 and M4 presents the opposite behavior to the diamagnetic models, with those nearer to
the black hole being more dense. Nevertheless, by comparison to the profiles of $\rho$ in Fig. \ref{variables_chi_cte}, we note that a nonconstant magnetic susceptibility does not affect the qualitative behavior of the rest mass density.

Now, from the right column of Fig. \ref{variables_chi_NOcte} it is clear that the gradients of $\chi_m$ determine the magnetization state of the disk, especially in the region between the inner edge and the center of the disk, where the spatial changes of $\chi_{m}$ are large. In this region, when $\partial\chi_{m}/\partial r<0$, as in models M1 and M3, the disks become more magnetized than in the Komissarov solution, while in the models M2 and M4, for which $\partial\chi_{m}/\partial r>0$, the disks considerably reduce their magnetization. Therefore, the stationary state of our tori implies that when the magnetic susceptibility decreases with the radial coordinate, the disk becomes more magnetized and vice versa.

\bgroup
\def\arraystretch{1.5}
\begin{table}
\begin{center}
\caption{\label{table1} Parameters defining the magnetic susceptibility of the disk in each model.}
\begin{tabular}{|c|c|c|c|c|c|}\hline \hline
 & M1 & M2 & A & M3 & M4 \\ \hline \hline
$\chi_{0}$ & -0.67 & -1$\times 10^{-5}$ & 0 & 1$\times 10^{-4}$ & 0.287 \\  \hline \hline
$\chi_{1}$ & 1.6 & -1.6 & 0 & 0.6 & -0.6 \\  \hline \hline
$\alpha$ & -1 & -1 & 1 & -1 & -1 \\  \hline \hline
\end{tabular}
\end{center}
\end{table} 
\egroup

\section{Conclusions}
\label{sec:conclusions}

In this paper we obtain for the first time a stationary and axisymmetric family of solutions for magnetized tori with magnetic polarization around a Kerr black hole. The magnetized tori was built by assuming a barotropic equation of state, a constant angular momentum, and a purely toroidal magnetic field. These features were previously considered by \cite{2006MNRAS.368..993K} to construct tori with dynamically important magnetic fields. In addition, the polarization of the tori was introduced by assuming the usual linear constitutive relation in which the magnetization vector and the applied magnetic field are linearly related by the magnetic susceptibility $\chi_{m}$. We show that our models reduce to the Komissarov solution when we set the magnetic susceptibility to zero, making it possible to determine the differences in the physical variables of a tori without magnetic polarization and a paramagnetic or diamagnetic tori.

With the aim of obtaining an analytical solution of the relativistic Euler equations, we assumed that $\chi=\chi(\mathcal{L}(r,\theta))$, and in particular, that $\chi$ takes the form (\ref{chi}). In this way, the polarization state of the tori is completely defined by choosing the constants $\alpha$, $\chi_{0}$, and $\chi_{1}$. In this paper we present two kinds of magnetized tori, one with constant magnetic susceptibility and one where this susceptibility varies. In the models of Sect. \ref{subsec:case1}, where $\chi_{m}$ is constant, we find that a paramagnetic torus ($\chi_{m}<0$) is more compact than a diamagnetic one ($\chi_{m}<0$), because the matter is more concentrated in the region between the inner edge and the center of the disk. However, in $r<r_c$  the diamagnetic tori are more magnetized than those obtained by Komissarov, and therefore, more magnetized than the paramagnetic ones. The opposite behavior for the magnetization parameter is obtained in the region $r>r_c$, where the paramagnetic tori are more magnetized than the case with $\chi_{m}=0$.

Finally, in the models of Sect. \ref{subsec:case2}, where the magnetic susceptibility is non-constant in the torus, we note that the rest mass density does not change its qualitative behavior, as compared with the models of constant $\chi_{m}$. Therefore, the way in which the matter is distributed in the tori depends on whether the fluid is diamagnetic or paramagnetic, the paramagnetic tori being more compact than the diamagnetic ones. Nevertheless, we find that the magnetization state of the tori also depends on the spatial changes of the magnetic susceptibility. In particular, when $\partial\chi_{m}/\partial r<0$ (models M1 and M3), the tori become more magnetized in $r<r_{c}$ than the Komissarov solution, and when $\partial\chi_{m}/\partial r>0$ (models M2 and M4), the tori lose considerable amounts of their magnetization in the same region of the disk. All these effects are more appreciable when the changes in the magnetic susceptibility are large, as in the inner region of the tori in our models. The next step in this direction is to add this solution as initial data, in the CAFE code \citep{2015ApJS..218...24L}, in order to carry out numerical simulations and see the dynamics of the accretion disk once it is being accreted onto the black hole. It is worth mentioning that in the last version of CAFE  the magnetic polarized matter terms were implemented following the characteristic approach presented in our recent work \citep{2018arXiv180602266P}.

\begin{acknowledgements}
O. M. P. would like to thank the financial support from COLCIENCIAS under the program Becas Doctorados Nacionales 647 and Universidad Industrial de Santander. F.D.L-C and G. A. G. were supported in part by VIE-UIS, under Grant No. 2314 and by COLCIENCIAS, Colombia, under Grant No. 8863.
\end{acknowledgements} 

%
%
%



\begin{thebibliography}{33}
\expandafter\ifx\csname natexlab\endcsname\relax\def\natexlab#1{#1}\fi

\bibitem[{{Abramowicz} {et~al.}(1978){Abramowicz}, {Jaroszynski}, \&
  {Sikora}}]{1978A&A....63..221A}
{Abramowicz}, M., {Jaroszynski}, M., \& {Sikora}, M. 1978, \aap, 63, 221

\bibitem[{{Abramowicz} \& {Fragile}(2013)}]{2013LRR....16....1A}
{Abramowicz}, M.~A. \& {Fragile}, P.~C. 2013, Living Reviews in Relativity, 16,
  1

\bibitem[{{Arajs} \& {Colvin}(1964)}]{1964JAP....35.2424A}
{Arajs}, S. \& {Colvin}, R.~V. 1964, Journal of Applied Physics, 35, 2424

\bibitem[{{Balbus} \& {Hawley}(1991)}]{1991ApJ...376..214B}
{Balbus}, S.~A. \& {Hawley}, J.~F. 1991, Astrophys. J., 376, 214

\bibitem[{{Balbus} \& {Hawley}(1998)}]{1998RvMP...70....1B}
{Balbus}, S.~A. \& {Hawley}, J.~F. 1998, Rev. Mod. Phys., 70, 1

\bibitem[{{Blandford} \& {Hernquist}(1982)}]{1982JPhC...15.6233B}
{Blandford}, R.~D. \& {Hernquist}, L. 1982, Journal of Physics C Solid State
  Physics, 15, 6233

\bibitem[{{Blandford} \& {Znajek}(1977)}]{1977MNRAS.179..433B}
{Blandford}, R.~D. \& {Znajek}, R.~L. 1977, Mon. Not. Roy. Astron. Soc., 179,
  433

\bibitem[{{Bugli} {et~al.}(2018){Bugli}, {Guilet}, {M{\"u}ller}, {Del Zanna},
  {Bucciantini}, \& {Montero}}]{2018MNRAS.475..108B}
{Bugli}, M., {Guilet}, J., {M{\"u}ller}, E., {et~al.} 2018, \mnras, 475, 108

\bibitem[{{Chatterjee} {et~al.}(2015){Chatterjee}, {Elghozi}, {Novak}, \&
  {Oertel}}]{2015MNRAS.447.3785C}
{Chatterjee}, D., {Elghozi}, T., {Novak}, J., \& {Oertel}, M. 2015, \mnras,
  447, 3785

\bibitem[{{Chevalier} {et~al.}(2002){Chevalier}, {Kahn}, {Bobet}, {Pasturel},
  \& {Etourneau}}]{2002JPCM...14L.365C}
{Chevalier}, B., {Kahn}, M.~L., {Bobet}, J.-L., {Pasturel}, M., \& {Etourneau},
  J. 2002, Journal of Physics Condensed Matter, 14, L365

\bibitem[{De~Haas \& Van~Alphen(1930)}]{de1930dependence}
De~Haas, W. \& Van~Alphen, P. 1930in , 170

\bibitem[{Dixon(1970)}]{dixon1970dynamics}
Dixon, W.~G. 1970, Proc. R. Soc. Lond. A, 314, 499

\bibitem[{{Font}(2008)}]{2008LRR....11....7F}
{Font}, J.~A. 2008, Living Reviews in Relativity, 11, 7

\bibitem[{{Fragile} \& {Meier}(2009)}]{2009ApJ...693..771F}
{Fragile}, P.~C. \& {Meier}, D.~L. 2009, \apj, 693, 771

\bibitem[{{Fragile} \& {S{\c a}dowski}(2017)}]{2017MNRAS.467.1838F}
{Fragile}, P.~C. \& {S{\c a}dowski}, A. 2017, \mnras, 467, 1838

\bibitem[{{Frank} {et~al.}(2002){Frank}, {King}, \&
  {Raine}}]{2002apa..book.....F}
{Frank}, J., {King}, A., \& {Raine}, D.~J. 2002, {Accretion Power in
  Astrophysics: Third Edition}, 398

\bibitem[{Gabold {et~al.}(2018)Gabold, Luan, Paul, Opel, M{\"u}ller-Buschbaum,
  Law, \& Paul}]{gabold2018structural}
Gabold, H., Luan, Z., Paul, N., {et~al.} 2018, Scientific reports, 8, 4835

\bibitem[{{Gimeno-Soler} \& {Font}(2017)}]{2017A&A...607A..68G}
{Gimeno-Soler}, S. \& {Font}, J.~A. 2017, \aap, 607, A68

\bibitem[{Griffiths(2005)}]{griffiths2005introduction}
Griffiths, D.~J. 2005, Introduction to electrodynamics

\bibitem[{{Huang} {et~al.}(2010){Huang}, {Huang}, {Rischke}, \&
  {Sedrakian}}]{2010PhRvD..81d5015H}
{Huang}, X.-G., {Huang}, M., {Rischke}, D.~H., \& {Sedrakian}, A. 2010, \prd,
  81, 045015

\bibitem[{{Komissarov}(2006)}]{2006MNRAS.368..993K}
{Komissarov}, S.~S. 2006, Mon. Not. Roy. Astron. Soc., 368, 993

\bibitem[{{Lora-Clavijo} {et~al.}(2015){Lora-Clavijo}, {Cruz-Osorio}, \&
  {Guzm{\'a}n}}]{2015ApJS..218...24L}
{Lora-Clavijo}, F.~D., {Cruz-Osorio}, A., \& {Guzm{\'a}n}, F.~S. 2015, \apjs,
  218, 24

\bibitem[{Lorrain \& Corson(1970)}]{citeulike:4033945}
Lorrain, P. \& Corson, D. 1970, Electromagnetic fields and waves (Freeman)

\bibitem[{Maugin(1978)}]{Maugin:1978tu}
Maugin, G.~A. 1978, J. Math. Phys., 19, 1198

\bibitem[{{McKinney} {et~al.}(2012){McKinney}, {Tchekhovskoy}, \&
  {Blandford}}]{2012MNRAS.423.3083M}
{McKinney}, J.~C., {Tchekhovskoy}, A., \& {Blandford}, R.~D. 2012, Mon. Not.
  Roy. Astron. Soc., 423, 3083

\bibitem[{Navarro-Noguera {et~al.}(2018)Navarro-Noguera, Lora-Clavijo, \&
  Gonz{\'a}lez}]{navarro2018general}
Navarro-Noguera, A., Lora-Clavijo, F., \& Gonz{\'a}lez, G.~A. 2018, General
  Relativity and Gravitation, 50, 76

\bibitem[{{Papapetrou}(1951)}]{1951RSPSA.209..248P}
{Papapetrou}, A. 1951, Proceedings of the Royal Society of London Series A,
  209, 248

\bibitem[{{Pimentel} {et~al.}(2018){Pimentel}, {Lora-Clavijo}, \&
  {Gonz{\'a}lez}}]{2018arXiv180602266P}
{Pimentel}, O.~M., {Lora-Clavijo}, F.~D., \& {Gonz{\'a}lez}, G.~A. 2018,
  Accepted for publication in ApJ [\eprint[arXiv]{1806.02266}]

\bibitem[{{Saijo} {et~al.}(1998){Saijo}, {Maeda}, {Shibata}, \&
  {Mino}}]{1998PhRvD..58f4005S}
{Saijo}, M., {Maeda}, K.-I., {Shibata}, M., \& {Mino}, Y. 1998, \prd, 58,
  064005

\bibitem[{{Semer{\'a}k}(1999)}]{1999MNRAS.308..863S}
{Semer{\'a}k}, O. 1999, \mnras, 308, 863

\bibitem[{{Suh} \& {Mathews}(2010)}]{2010ApJ...717..843S}
{Suh}, I.-S. \& {Mathews}, G.~J. 2010, Astrophys. J., 717, 843

\bibitem[{{Wang} {et~al.}(2016){Wang}, {L{\"u}}, {Zhu}, \&
  {Wu}}]{2016PASP..128j4201W}
{Wang}, Z., {L{\"u}}, G., {Zhu}, C., \& {Wu}, B. 2016, \pasp, 128, 104201

\bibitem[{{Wielgus} {et~al.}(2015){Wielgus}, {Fragile}, {Wang}, \&
  {Wilson}}]{2015MNRAS.447.3593W}
{Wielgus}, M., {Fragile}, P.~C., {Wang}, Z., \& {Wilson}, J. 2015, \mnras, 447,
  3593

\end{thebibliography}

\end{document}